\documentclass[12pt]{article}
\pdfoutput=1
\usepackage{geometry,enumerate,amsmath,amssymb}
\usepackage{fullpage}
\usepackage{graphicx}
\usepackage{bm}
\numberwithin{equation}{section}
\newcommand{\be}{\begin{equation}}
\newcommand{\ee}{\end{equation}}
\newcommand{\bea}{\begin{eqnarray}}
\newcommand{\eea}{\end{eqnarray}}

\renewcommand{\hat}{\widehat}

\renewcommand{\epsilon}{\varepsilon}
\newcommand{\sech}{\mbox{sech}\,}

\begin{document}
\title{
Boundary metrics on soliton moduli spaces
}
\author{
  Paul Sutcliffe\\[10pt]
 {\em \normalsize Department of Mathematical Sciences,}\\
 {\em \normalsize Durham University, Durham DH1 3LE, United Kingdom.}\\ 
{\normalsize Email:  p.m.sutcliffe@durham.ac.uk}
}
\date{October 2021}

\maketitle
\begin{abstract}
  The geodesic approximation is a powerful method for studying the dynamics of BPS solitons. However, there are systems, such as BPS monopoles in three-dimensional hyperbolic space, where this approach is not applicable because the moduli space metric defined by the kinetic energy is not finite. In the case of hyperbolic monopoles, an alternative metric has been defined using the abelian connection on the sphere at infinity, but its relation to the dynamics of hyperbolic monopoles is unclear. Here this metric is placed in a more general context of boundary metrics on soliton moduli spaces. Examples are studied in systems in one and two space dimensions, where it is much easier to compare the results with simulations of the full nonlinear field theory dynamics. It is found that geodesics of the boundary metric provide a reasonable description of soliton dynamics.
\end{abstract}

\newpage
\section{Introduction}\quad
In the geodesic approximation the motion of slowly moving BPS solitons is described by geodesic motion in the moduli space of static solitons, equipped with a metric induced by the kinetic energy of the field theory \cite{Ma1}. This approach has been applied to a wide variety of soliton systems, following its initial application to the study of magnetic monopole dynamics in three-dimensional Euclidean space \cite{AH}. The physical intuition underlying the success of this approximation is that slowly moving solitons possess only a small amount of kinetic energy and therefore energy conservation prevents the field configuration from deforming far from the moduli space of minimal energy static soliton solutions. Projecting the dynamics to an adiabatic evolution in the moduli space therefore appears a reasonable simplification. This physical argument is supported by rigorous mathematical justification in the case of vortices and monopoles \cite{Stu2,Stu}.

There are some BPS soliton systems where the geodesic approximation is not applicable because restricting the dynamics to motion in the moduli space yields infinite kinetic energy, and therefore is not a prescription to obtain a metric on the soliton moduli space. A notable example is the case of BPS monopoles in three-dimensional hyperbolic space. The static properties of monopoles in hyperbolic space share many of the features of monopoles in Euclidean space \cite{At}, but this crucial difference in the divergence of the metric means that nothing is known about the dynamics of hyperbolic monopoles.
Braam and Austin defined a metric on the moduli space of hyperbolic monopoles using the abelian connection on the sphere at infinity \cite{BA}, but its relation to the dynamics of hyperbolic monopoles remains unclear. The moduli space of a single hyperbolic monopole equipped with the Braam-Austin metric is three-dimensional hyperbolic space, which is encouraging, as interpreting its geodesics as the free motion of a single hyperbolic monopole seems reasonable. Simulating the full field dynamics of hyperbolic monopoles would be a worthy endeavour, although this is by no means an easy task.

In this paper an alternative approach is adopted to gain some support for the relevance of the Braam-Austin metric to the dynamics of hyperbolic monopoles. This metric is placed in a more general context of boundary metrics on BPS soliton moduli spaces. The metric is obtained by a simple renormalization of the divergent metric induced from the field theory kinetic energy. By introducing a cutoff, the leading divergence can be identified and used to renormalize the metric via multiplication by a factor that depends only on the cutoff. Multiplication by this factor, of course, does not change the geodesic equation, so the basic fact being exploited by this approach is that there is a convergence for the geodesic equation as the cutoff is removed, even though the natural metric diverges. The geodesic equation obtained in the limit is the equation for the geodesics of the renormalized metric, which is termed a boundary metric because it depends only the soliton fields at spatial infinity. The next section provides the details of the construction of the boundary metric for hyperbolic monopoles, and further sections consider a radial lump in the $\mathbb{CP}^1$ sigma model in the plane, and kink dynamics on a line. The advantage of low-dimensional examples is that comparisons with full nonlinear field theory simulations are much easier to investigate.

\section{Hyperbolic monopoles}\quad
This section concerns $SU(2)$ magnetic monopoles in three-dimensional hyperbolic space of curvature $-1$, with metric
\be
ds^2({\mathbb H}^3)=\frac{4(dx_1^2+dx_2^2+dx_3^2)}{(1-r^2)^2},
\ee
where $r^2=x_1^2+x_2^2+x_3^2<1.$ The Higgs field $\Phi$ and the gauge potential $A_i$ take values in $\mathfrak{su}(2)$, and the monopole scale relative to the curvature of hyperbolic space is fixed by the boundary condition
$|\Phi|^2=-\frac{1}{2}\mbox{tr}(\Phi^2)\to\frac{1}{4}$ as $r\to 1.$

Hyperbolic monopoles are solutions of the Bogomolny equation
\be
F_{ij}=\frac{2}{1-r^2}\varepsilon_{ijk} D_k\Phi,
\label{hmon_bog}
\ee
where $\varepsilon_{ijk}$ is the totally anti-symmetric tensor, 
$F_{ij}=\partial_i A_j-\partial_j A_i+[A_i,A_j]$ is the gauge field, and
$D_i\Phi=\partial_i\Phi+[A_i,\Phi]$ is the covariant derivative of the Higgs field.

The Higgs field at infinity defines a map between two-spheres, 
$\Phi|_{r=1}:S^2\to S^2$, and the monopole charge $N$ is the associated positive integer element of the homotopy group  $\pi_2(S^2)={\mathbb Z}.$
The moduli space, ${\cal M}_N$, of charge $N$ monopole solutions of the Bogomolny equation (\ref{hmon_bog}), up to gauge transformations, is a $(4N-1)$-dimensional manifold \cite{At}. In regions of this moduli space where the monopoles are well-separated, these parameters may be interpreted as a position for each monopole and a set of relative phases.

Monopole solutions of the Bogomolny equation are static solutions of a dynamical $SU(2)$ Yang-Mills-Higgs theory with metric $ds^2=dt^2-ds^2({\mathbb H}^3)$ and kinetic energy
\be
T=\int\limits_{r\le 1}
-\mbox{tr}\bigg\{
  \frac{2}{(1-r^2)^3}(D_t\Phi)^2+\frac{1}{2(1-r^2)}E_i^2
  \bigg\}dx_1dx_2dx_3,
  \label{hmon_kinetic}
  \ee
  where $D_t\Phi=\dot\Phi+[A_t,\Phi]$ and $E_i=\dot A_i-\partial_i A_t+[A_t,A_i],$ with a dot denoting differentiation with respect to $t.$
The equation of motion for $A_t$ that follows from the variation of (\ref{hmon_kinetic}) is Gauss' law
  \be
  D_i\bigg(\frac{E_i}{1-r^2}\bigg)=\frac{4}{(1-r^2)^3}[D_t\Phi,\Phi].
  \ee
  In the gauge $A_t=0$ the kinetic energy simplifies to
  \be
T=\int\limits_{r\le 1}
-\mbox{tr}\bigg\{
  \frac{2}{(1-r^2)^3}\dot\Phi^2+\frac{1}{2(1-r^2)}\dot A_i^2
  \bigg\}d^3x,
  \label{hmon_kinetic2}
  \ee
  subject to the constraint from Gauss' law 
  \be
  D_i\bigg(\frac{\dot A_i}{1-r^2}\bigg)=\frac{4}{(1-r^2)^3}[\dot\Phi,\Phi],
  \label{gauss2}
  \ee
  to ensure that the motion is orthogonal to the gauge orbit.

  In the geodesic approximation the time evolution of the fields is restricted to time evolution in the moduli space. Taking the time derivative of (\ref{hmon_bog}) yields
  \be
  D_i\dot A_j-D_j \dot A_i =
  \frac{2}{1-r^2}\varepsilon_{ijk} (D_k\dot\Phi+[\dot A_k,\Phi]).
  \label{hmon_linbog}
  \ee
  The solutions $(\dot\Phi,\dot A_i)$ of (\ref{hmon_linbog}) and (\ref{gauss2}) are the tangent vectors, but each has a length-squared, given by the kinetic energy (\ref{hmon_kinetic2}), that is infinite. This is the reason that the geodesic approximation is not applicable to hyperbolic monopoles.
  To examine the details of this divergence let
  $T_b$ denote the regularized kinetic energy evaluated with a cutoff, so that the region of integration is $r\le b$, with $T_b$ finite for $0<b<1.$
  A naive examination of (\ref{hmon_kinetic2}) might conclude that the first term leads to a divergence as $b\to 1,$ due to the appearance of the factor $(1-r^2)^{-3}.$ However, the condition of orthogonality to gauge orbits yields a finite limit for $\dot\Phi(1-r^2)^{-2}$ as $r\to 1$, so there is no divergence of the kinetic energy as $b\to 1$ from this first term. The divergence arises solely from the contribution of $\dot A_i$ in the direction of $\Phi,$ because the abelian component $\dot a_i=-\mbox{tr}(\dot A_i\Phi)$ is finite as $r\to 1.$ This produces a logarithmic divergence with $-T_b/\log(1-b)$ finite as $b\to 1.$

  The boundary metric is defined by the associated renormalized kinetic energy
  \be
  \hat T= -\lim_{b\to 1} T_b/\log(1-b)
  =\int\limits_{S^2} \frac{1}{2}\dot a_i^2 \sin\theta\,d\theta d\varphi,
  \ee
  where the integration is over the boundary sphere $r=1$, with $\theta$ and $\varphi$ the usual angular coordinates on $S^2.$ On this boundary sphere Gauss' law (\ref{gauss2}) projected onto the abelian component reduces to $a_r=0$ and the
  requirement that the angular components must satisfy
  \be
  \partial_\theta(\dot a_\theta \sin\theta)+\frac{\partial_\varphi \dot a_\varphi}{\sin\theta}=0.
  \label{gauss3}
  \ee
  Finally, with this condition, the renormalized kinetic energy on the monopole moduli space becomes that of an abelian gauge theory on the sphere
  \be
  \hat T
  =\int\limits_{S^2} \frac{1}{2}\bigg(\dot a_\theta^2 +\frac{\dot a_\varphi^2}{\sin^2\theta}\bigg)\sin\theta\,d\theta d\varphi.
  \label{kinetic_s2}
  \ee
  The fact that this defines a metric on the monopole moduli space relies on the result that, up to gauge transformations, the abelian connection $a=a_\theta d\theta+a_\varphi d\varphi$ on $S^2$ determines the monopole \cite{BA}.

  The boundary metric defined by the renormalized kinetic energy (\ref{kinetic_s2}) is the metric introduced by Braam and Austin \cite{BA}. They used a symmetry argument to show that the single monopole moduli space, ${\cal M}_1$, equipped with this metric, is three-dimensional hyperbolic space ${\mathbb H}^3$. More explicitly, let ${\bf X}$ be the point in the interior of the unit ball that corresponds to the position of the monopole, given by the location at which $\Phi$ vanishes.
  In a suitable gauge
  \be
  \dot a_\theta=-\mbox{tr}\bigg(\Phi \dot {\bf X} \cdot \frac{\partial A_\theta}{\partial {\bf X}}\bigg)\bigg|_{r=1}, \qquad 
   \dot a_\varphi=-\mbox{tr}\bigg(\Phi \dot {\bf X} \cdot \frac{\partial A_\varphi}{\partial {\bf X}}\bigg)\bigg|_{r=1},
   \ee
   satisfy (\ref{gauss3}) and then (\ref{kinetic_s2}) yields
  \be
  \hat T =\frac{4\pi}{3}\frac{|\dot {\bf X}|^2}{(1-|{\bf X}|^2)^2},
  \ee
  which is indeed proportional to $ds^2({\mathbb H}^3).$

  To even consider the relevance to hyperbolic monopole dynamics of the geodesics of the boundary metric given by (\ref{kinetic_s2}), it is necessary to clarify the type of field theory initial conditions that might be appropriate. Clearly, the fields $(\Phi,A_i)$ at an initial time should be taken to be those of a point in the moduli space ${\cal M}_N$, but the initial time derivatives cannot be taken to be a tangent vector $(\dot\Phi,\dot A_i)$, as this requires infinite kinetic energy and is therefore unphysical. The suggestion here is to curtail the tangent vector by multiplying it by a smoothed indicator function $\chi$, satisfying  $\chi\approx 1$ in the core of the monopoles and $\chi\approx 0$ far from the core of any monopole, with $\chi\to 0$ as $r\to 1$ sufficiently rapidly so that the kinetic energy is finite. A suitable choice is $\chi = 1-(2|\Phi|)^m$, where $m$ is a large positive integer. For any initial condition of this form, causality considerations imply that for a sufficiently large $m$, the evolution within any given spatial region will be insensitive to the value of $m$ for evolution up to some time limit that can be increased by increasing $m$.

  \section{Radial lumps in the $\mathbb{CP}^1$ sigma model}\quad
  For a second example of a boundary metric, consider the $\mathbb{CP}^1$ sigma model, in three-dimensional Minkowski spacetime, given by the Lagrangian density
  \be
     {\cal L}=\frac{\partial_\mu W \partial^\mu \overline{W}}{(1+|W|^2)^2}.
     \ee
     Here $W$ takes values in the Riemann sphere and satisfies the boundary condition $W\to 0$ as $x_1^2+x_2^2\to\infty.$ This one-point compactification of the plane implies that at any fixed time $W:{\mathbb R}^2\cup\{\infty\}\mapsto \mathbb {CP}^1$, with an associated topological charge $N\in{\mathbb Z}=\pi_2(\mathbb {CP}^1)$
     that counts the number of solitons, often known as lumps in this context.

For each positive integer $N$, the general static $N$-lump solution is given by taking $W$ to be a rational function of $x_1+ix_2$ with degree $N$. Restricting attention to static radially symmetric charge $N$ lumps positioned at the origin gives a two-dimensional moduli space of solutions 
\be
     W=\frac{\lambda^N e^{i\psi}}{(x_1+ix_2)^N},
     \ee
     with $\lambda>0$ a measure of the size of the lump and $\psi$ an internal phase.

     For $N>1$ the geodesic approximation is applicable to motion in this two-dimensional moduli space as the kinetic energy is finite. Explicitly, allowing $\lambda$ and $\psi$ to be time-dependent produces the kinetic energy
     \be
     T=\int \frac{|\dot W|^2}{(1+|W|^2)^2}\, d^2x
     =(N^2\dot\lambda^2+\lambda^2\dot\psi^2)\frac{\pi^2}{N^2\sin(\pi/N)},
     \ee
     yielding a flat metric on this moduli space.
     However, for $N=1$ there is a logarithmic divergence in the kinetic energy for motion in the moduli space, making it a relevant example for the study of boundary metrics. As in the previous section, the boundary metric is defined by the associated renormalized kinetic energy $\hat T$.
Introducing polar coordinates $r,\theta$ in the plane, let $T_b$ denote the regularized kinetic energy defined by integration over the disc $r\le b$, and set  
       \be
       \hat T=\lim_{b\to\infty}\frac{T_b}{\log b}
       =\int_0^{2\pi} \frac{|\dot W|^2r^2}{(1+|W|^2)^2}\bigg|_{r=\infty} d\theta
       \ =(\dot\lambda^2+\lambda^2\dot\psi^2)2\pi,
       \ee
       where the integration is over the boundary circle at infinity.

Note that the geodesics of this boundary metric fit naturally as the $N=1$ member of the geodesics of the general family of flat metrics
       \be
       ds^2=N^2d\lambda^2+\lambda^2d\psi^2,
       \ee
       proportional to those obtained from the geodesic approximation with $N>1.$ For all $N$ the qualitative dynamics is the same. If the internal phase is constant, $\dot \psi=0$, then the size $\lambda$ of the lump either increases without limit, or shrinks to zero in a finite time, as this family of metrics are geodesically incomplete. If there is motion in the internal phase then the non-zero conserved quantity $\lambda^2\dot\psi$ prevents the lump from shrinking to zero size and any initial shrinking is eventually reversed and ultimately the lump expands without limit.

       The qualitative features predicted by the geodesic approximation using the boundary metric are in good agreement with the results from field theory simulations \cite{LPZ1,PZ1}. Numerical studies of the shrinking to zero size in finite time for a $1$-lump confirm that a linear decrease in the size of the lump remains an accurate description right up until times very close to the critical time at which the size vanishes. The simulations  are performed on a numerical grid with a finite spatial extent, taken large enough so that causality prevents any influence of the boundary on the field at the centre of the lump during the simulation time \cite{PZ1}. This is equivalent to using an indicator function with $\chi=1$ in the simulation region, with the value outside this region being irrelevant for the evolution of the field at the centre of the lump, within the limited time of the simulation.

       The main difference between the field theory evolution and the geodesic approximation is attributed to the fact that the latter neglects radiation, which is essential in expelling excess energy from the core of the lump \cite{BCT}. This applies equally well to all values of $N$, but for $N>1$ an initial slow evolution tangential to the moduli space implies that the kinetic energy is small compared to the potential energy, whereas this is not the case for $N=1$. The geodesic approximation using the boundary metric models field theory evolution in which the kinetic energy can be arbitrarily large, depending upon the details of the indicator function $\chi,$ or equivalently the spatial extent of the simulation region in numerical simulations.

A radially symmetric lump located at the origin is an appropriate system for the application of the geodesic approximation using the boundary metric, because all tangent vectors in the two-dimensional moduli space have infinite length. This is complementary to previous studies of lump dynamics using the geodesic approximation restricted to only those tangent vectors that have finite length \cite{Wa3,Le2}. Neither approach is applicable for motion in the full moduli space of $N$-lumps, because of the existence of tangent vectors of both finite and infinite length.         
     
\section{Kink dynamics}\quad
The simplest system to illustrate and investigate a boundary metric on a soliton moduli space is a kink in one spatial dimension. Consider a modified $\phi^4$ theory given by the Lagrangian density 
\be
   {\cal L}=\frac{1}{2}\dot\phi^2 \cosh(\kappa x) -\frac{1}{2}\phi'^2 -\frac{1}{2}(1-\phi^2)^2,
   \ee
   where $\kappa$ is a non-negative constant and prime denotes differentiation with respect to the spatial coordinate $x.$ The standard $\phi^4$ theory is obtained for $\kappa=0.$ As the static sector is independent of $\kappa$ then there is a one-dimensional moduli space of static kink solutions given by the usual $\phi^4$ kink 
   \be \phi(x,t)=\tanh(x-a), \ee
  where the arbitrary real constant $a$ is a coordinate on the moduli space corresponding to the position of the kink.

\begin{figure}[ht]\begin{center}
    \includegraphics[width=0.5\columnwidth]{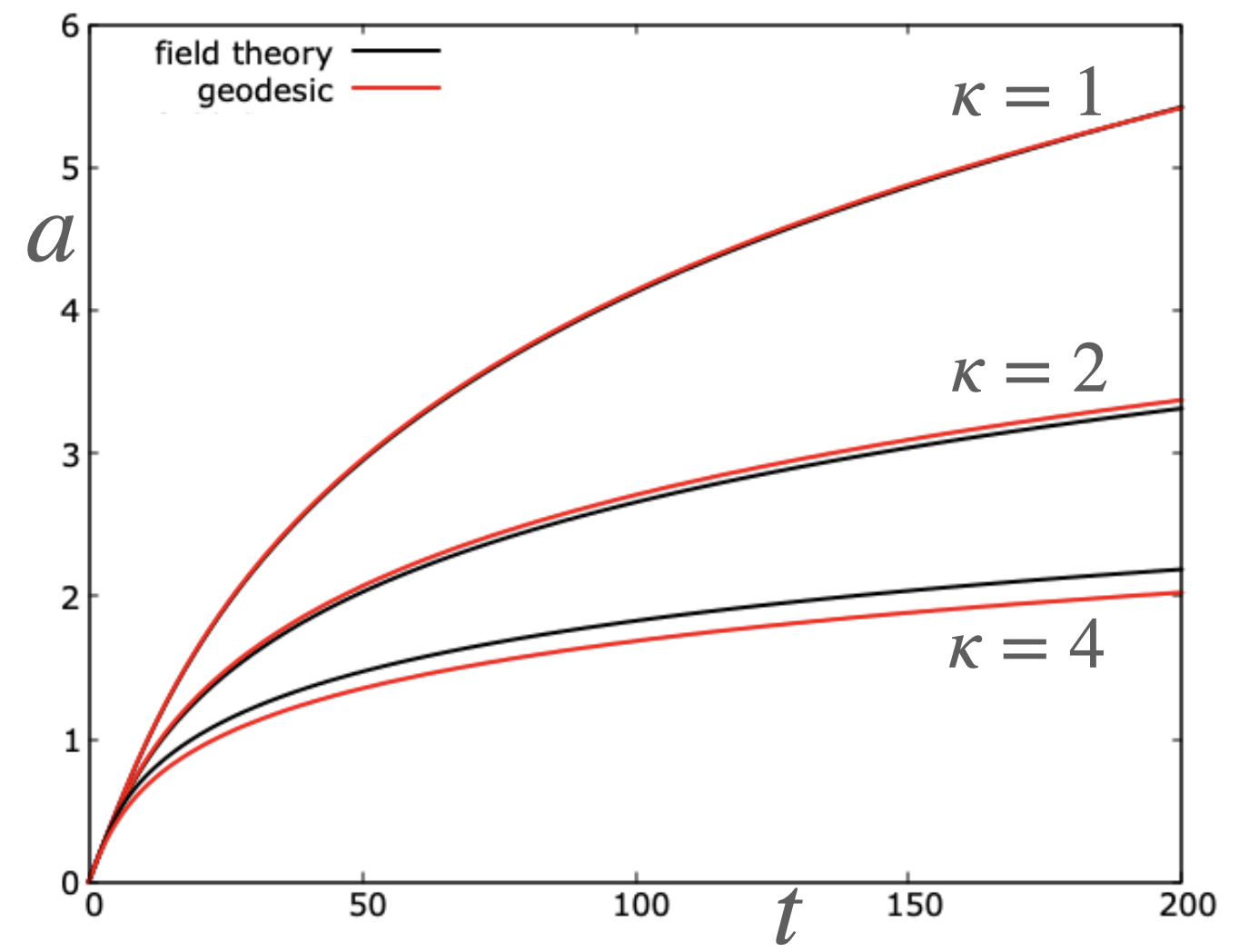}
\caption{The black curves give the position $a$ of the kink obtained from a field theory simulation with an initial condition given by $a_0=0$ and $v=0.1.$ The red curves are the corresponding geodesic approximation, using the boundary metric in the case $\kappa=4$.}
\label{fig1}\end{center}\end{figure}

The geodesic approximation restricts the motion to the moduli space by promoting $a$ to be time dependent, producing the tangent vector  
   \be \dot\phi=-\dot a \,\sech^2(x-a). \ee
   If $0\le\kappa<4$ then the associated kinetic energy 
   \be T=\int_{-\infty}^\infty \frac{1}{2}\dot\phi^2 \cosh(\kappa x)\,dx
   \label{kinkke}
   \ee
   is finite and the standard geodesic approximation is applicable.
   For example, if $\kappa=1$ then $T=\frac{\pi}{4}\dot a^2 \cosh a$ with geodesics given by $\ddot a=-\frac{1}{2} \dot a^2 \tanh a.$
   The upper red curve in Fig.\ref{fig1} shows the geodesic with $a(0)=a_0=0$ and $\dot a(0)=v=0.1.$

It is a simple task to compare the results of the geodesic approximation with field theory simulations using the initial condition
   \be \phi(x,0)=\tanh(x-a_0), \quad \dot \phi(x,0)=-v\,\sech^2(x-a_0). \ee
The results presented below use a numerical scheme with 6001 lattice points and a lattice spacing $\Delta x=0.0025$ giving a simulation interval $[-7.5,7.5]$. Spatial derivatives are approximated using fourth-order accurate finite differences and time evolution is performed using a fourth-order Runge-Kutta method with a timestep $\Delta t=\frac{1}{2}\Delta x.$  No flux boundary conditions are imposed at the ends of the interval. The position of the kink, $a$, is extracted using linear interpolation between the two lattice points at which $\phi$ changes sign. In Fig.\ref{fig1} there is in fact an upper black curve showing the results of the field theory simulation, but it is almost totally obscured by the red curve, because the geodesic approximation is so accurate in this case.

To see a slight difference between the geodesic approximation and field theory simulations, consider the case $\kappa=2.$ The kinetic energy is $T=\frac{4}{3}\dot a^2 \cosh(2a)$ with geodesics given by $\ddot a=-\dot a^2 \tanh(2a).$
The middle red curve in Fig.\ref{fig1} again shows the geodesic with $a_0=0$ and $v=0.1,$ and the black curve is the field theory result that displays just enough difference to allow separate curves to be distinguished.

Turning now to the case $\kappa=4$, then the kinetic energy is infinite if motion is restricted to the moduli space, making it a suitable example for the application of a boundary metric. Let $T_b$ denote the regularized kinetic energy integrated over the interval $[-b,b].$  As $T_b$ is linearly divergent as $b\to\infty,$ define the renormalized kinetic energy
\be
\hat T= \lim_{b\to\infty} \frac{T_b}{b}
=\lim_{x\to \infty}\bigg[\frac{1}{2}\dot\phi^2\cosh(4x)\bigg]
+\lim_{x\to -\infty}\bigg[\frac{1}{2}\dot\phi^2\cosh(4x)\bigg]
=8\dot a^2\cosh(4a).
\ee
The geodesics of this boundary metric are given by
$\ddot a=-2\dot a^2 \tanh(4a).$
The lower red curve in Fig.\ref{fig1} again shows the geodesic with $a_0=0$ and $v=0.1,$ and the black curve is the field theory result. This demonstrates that geodesics of the boundary metric provide a reasonable description of kink dynamics. As earlier, the field theory computation should be regarded as equivalent to using an indicator function with $\chi=1$ in the simulation region, with the value outside this region being irrelevant for the evolution of the kink over the timescale of the simulation. This has been justified by performing the same computation using double the spatial region and confirming that there is no discernible difference in the evolution of the position of the kink. Changing the boundary conditions from no flux to Dirichlet also yields no perceptible change, providing another numerical check.

\begin{figure}[ht]\begin{center}
 \hbox{   \includegraphics[width=0.5\columnwidth]{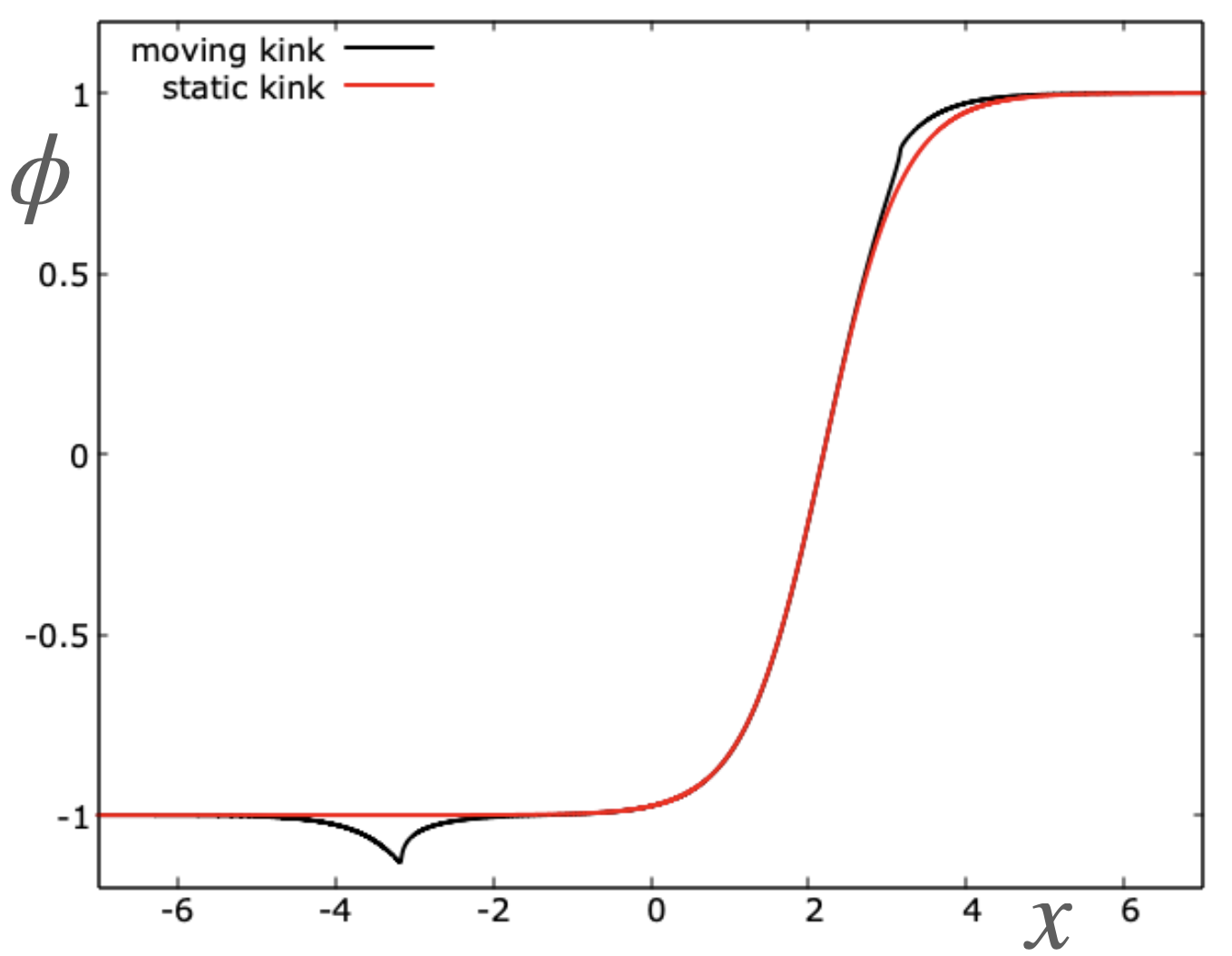}
        \includegraphics[width=0.5\columnwidth]{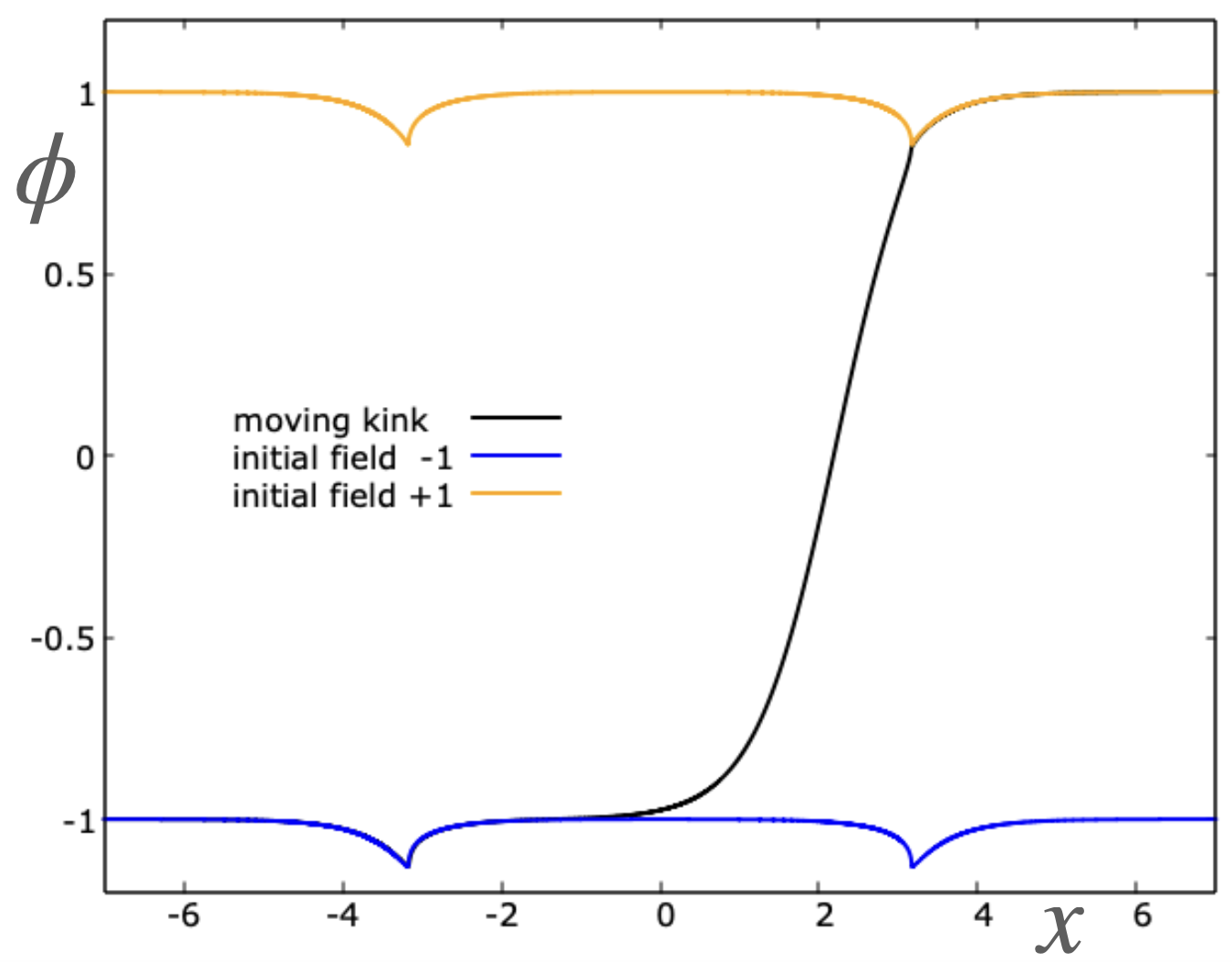}}
 \caption{Left: The moving kink (black curve) at $t=200$ and the static kink (red curve) with the same position.
   Right: The moving kink (black curve) at $t=200$. The field at $t=200$ from the initial condition $\phi=-1$ (blue curve) and $\phi=1$ (orange curve) with the initial time derivative $\dot\phi$ the same as for the moving kink.}
 \label{fig2}\end{center}\end{figure}

The expectation is that radiation is the main source of the discrepancy between field theory evolution and the geodesic approximation. In theories with a boundary metric, the slow motion of a soliton no longer implies that the kinetic energy is a small fraction of the total energy. A moving soliton may therefore generate a considerable amount of radiation and energy conservation does not provide such a useful constraint to prevent the field configuration from deforming far from the moduli space.
To illustrate this issue, the black curve in the left-hand-side plot in Fig.\ref{fig2} shows the field $\phi$ at the end of the simulation ($t=200$) associated with the data from Fig.\ref{fig1}. Also shown (red curve) is the static kink with the same position. This reveals that the two fields agree in the core of the kink, but that there are significant deformations in the regions around $x\approx \pm 3.$ To check that these deformations are consistent with an initial burst of radiation leaving the core of the kink, consider the equation for the characteristics that pass through the origin at $t=0$
\be
\int_0^x \sqrt{\cosh(4X)}\,dX=\pm t.
\ee
Evaluating this equation at $t=200$ yields $x=\pm 3.17$, which is indeed consistent with the location of the deformation.

Further evidence for the nature of the deformation is provided by considering initial conditions given by the vacuum fields $\phi(x,0)=-1$ and $\phi(x,0)=1$, with the initial time derivative $\dot\phi(x,0)$ the same as that of the moving kink. The results are displayed in the right-hand-side plot in Fig.\ref{fig2}, where the field at time $t=200$ is shown in blue for $\phi(x,0)=-1$
and orange for $\phi(x,0)=1$. For comparison, the moving kink field is displayed again as the black curve. This clearly shows that the moving kink may be decomposed into a soliton component, well-approximated by a point in the moduli space, and a substantial radiation component, which nonetheless has little influence on the soliton core.

\section{Conclusion}\quad
The concept of a boundary metric allows the geodesic approximation of soliton dynamics to be applied to situations in which motion in the soliton moduli space yields infinite kinetic energy. Numerical studies of a low-dimensional example provide support for this approach and suggest that it provides a reasonable description of the soliton component of the dynamics, even though there is more radiation than in the usual application of the geodesic approximation. This work provides motivation for a future study of the Braam-Austin metric on the moduli space of hyperbolic monopoles and an interpretation of the results in terms of hyperbolic monopole scattering. To date, only the metric on the moduli space of a single hyperbolic monopole has been calculated. Numerical field theory studies of hyperbolic monopole dynamics would also be interesting, for comparison to predictions from the geodesic approximation, although this is a significant numerical challenge.

It would be interesting to see if any rigorous mathematical results can be derived for the geodesic approximation with a boundary metric, along the lines of the results obtained for the usual geodesic approximation \cite{Stu2,Stu}.


\begin{thebibliography}{99}

\bibitem{Ma1} N.~S. Manton, 
A remark on the scattering of BPS monopoles,
\textit{Phys. Lett.} \textbf{B110}, 54 (1982).

\bibitem{AH} M.~F. Atiyah and N.~J. Hitchin,
\textit{The Geometry and Dynamics of Magnetic Monopoles},
Princeton University Press, 1988.

\bibitem{Stu2} D. Stuart, 
Dynamics of abelian Higgs vortices in the near Bogomolny regime,
\textit{Commun. Math. Phys.} \textbf{159}, 51 (1994).

\bibitem{Stu} D. Stuart, 
The geodesic approximation for the Yang-Mills-Higgs equations,
\textit{Commun. Math. Phys.} \textbf{166}, 149 (1994).
  
\bibitem{At}
M.~F. Atiyah, {Magnetic monopoles in hyperbolic spaces},
in \textit{M. Atiyah: Collected Works, vol. 5},
Oxford, Clarendon Press, 1988.

\bibitem{BA} P.~J. Braam and D.~M. Austin,
Boundary values of hyperbolic monopoles,
\textit{Nonlinearity} \textbf{3}, 809 (1990).

\bibitem{LPZ1} R.~A. Leese, M. Peyrard and W.~J. Zakrzewski,
Soliton stability in the $O(3)$ $\sigma$-model in (2+1) dimensions,
\textit{Nonlinearity} \textbf{3}, 387 (1990).

\bibitem{PZ1} B. Piette and W.~J. Zakrzewski,
Shrinking of solitons in the (2+1)-dimensional $S^2$  sigma model,
\textit{Nonlinearity} \textbf{9}, 897 (1996).

\bibitem{BCT} P. Bizo\'{n}, T. Chmaj and Z. Tabor,
Formation of singularities for equivariant $(2+1)$-dimensional wave maps into the 2-sphere,
\textit{Nonlinearity} \textbf{14}, 1041 (2001).

\bibitem{Wa3} R.~S. Ward, 
Slowly-moving lumps in the $\mathbb{CP}^1$ model in (2+1) dimensions,
\textit{Phys. Lett.} \textbf{B158}, 424 (1985).

\bibitem{Le2} R.~A. Leese, 
Low energy scattering of solitons in the $\mathbb{CP}^1$ model,
\textit{Nucl. Phys.} \textbf{B344}, 33 (1990).

\end{thebibliography}
\end{document}